\documentclass[prb,twocolumn,showpacs,amsmath,amssymb,floatfix,showpacs]{revtex4}
\usepackage{color,graphics,epsfig,rotating}
\usepackage{graphicx}
\usepackage{dcolumn}
\usepackage{bm}
\usepackage{subfigure}
\usepackage{ulem}

\begin{document}

\title{Electron-phonon superconductivity in $A$Pt$_3$P ($A$ = Sr, Ca, La) compounds: From weak to strong coupling}

\author{Alaska Subedi}
\affiliation{Max Planck Institute for Solid State Research, Heisenbergstr. 1, D-70569
Stuttgart, Germany}

\author{Luciano Ortenzi}
\affiliation{Max Planck Institute for Solid State Research, Heisenbergstr. 1, D-70569
Stuttgart, Germany}

\author{Lilia Boeri}
\affiliation{Max Planck Institute for Solid State Research, Heisenbergstr. 1, D-70569
Stuttgart, Germany}

\begin{abstract}
We study the newly-discovered Pt phosphides $A$Pt$_3$P ($A$=Sr, Ca, La)
[\textit{T. Takayama, et al.,
  Phys. Rev. Lett. {\bf 108}, 237001}] using first-principles
calculations and Migdal-Eliashberg theory.  Given the remarkable
agreement with the experiment, we exclude the charge-density wave
scenario proposed by previous first-principles calculations, and give
conclusive answers concerning the superconducting state in these
materials. The pairing increases from La to Ca and Sr due to changes
in the electron-phonon matrix elements and low frequency phonons.  Although we find that all three
compounds are well described by conventional $s$-wave
superconductivity and spin orbit coupling of Pt plays a marginal role,
we show that it could be possible to tune the structure from
centrosymmetric to non centrosymmetric opening new perspectives
towards the understanding of unconventional superconductivity.
\end{abstract}

 \pacs{74.70.-b, 74.20.Pq, 63-20.kd } 


\maketitle

\section{Introduction}

In the last ten years,
several important discoveries have sensibly 
advanced our understanding of superconductivity:
a record $T_c$ of 39 K in the BCS superconductor MgB$_2$,\cite{nagamatsu} exotic
superconductivity with $T_c$ of up to 56 K in the iron-based
superconductors,\cite{hosono} as well as superconductivity in
boron-doped diamond,\cite{diamond} aromatic compounds,\cite{picene}
and so on.
%
%
At the same time, ideas percolating from other fields of condensed
matter have brought in new twists into this old and fascinating
phenomenon. For example, the recent interest in spin-orbit coupling
(SOC)~\cite{SOC} has revived the discussion on superconductivity
in non-centrosymmetric crystals (NCSC),\cite{sigri:RMP,fisch11} %
boosted 
by the discovery of the heavy fermion CePt$_3$Si
($T_c$=0.75 K) in 2004.\cite{cpsi:finding}
%
In crystals without inversion symmetry, a  
strong antisymmetric SOC 
that lifts the spin
degeneracy can be conducive to
exotic pairing symmetry. 
Because of the large SOC of Pt ($Z = 78$),
this makes Pt-based compounds promising candidates for exotic
superconductivity, as discussed 
for SrPtAs.\cite{nish11:srptas,srptas:theory}

Recently, Takayama \textit{et al.} discovered 
a new family of ternary platinum phosphide superconductors with
chemical formula $A$Pt$_3$P ($A$ = Sr, Ca, and
La) and $T_c$'s of 8.4 K, 6.6 K and 1.5 K, respectively.\cite{taka12}
Besides the relatively high $T_c$,
these compounds exhibit 
a very interesting crystal structure, which is a 
centrosymmetric variant of the CePt$_3$Si one.
The authors have suggested 
that 
this discovery would have a very 
strong impact in the field of superconductivity
if one could
synthesize both centrosymmetric and non-centrosymmetric
variants of superconductors consisting of the electronically
  equivalent elements.
This would in fact
allow to study the effect of inversion symmetry on superconductivity
in a controlled way.
The  nature of 
the superconducting pairing in the $A$Pt$_3$P compounds has
been debated through
experiments~\cite{taka12} and \textit{ab initio}
calculations.\cite{nekr12,chen12,kang_cm}
 In the original
discovery paper, it was proposed
that, at least in SrPt$_3$P,
the superconductivity is of strong-coupling $s$-wave type
with clear signatures of low-lying phonons 
 and large BCS ratios suggestive of multiband behavior.
%
%
%
Ref.~\onlinecite{chen12}
has instead proposed 
that $T_c$
is enhanced by 
the proximity to a dynamical charge-density wave (CDW) instability,
and that a strong SOC could eventually lead to
\textit{exotic} superconductivity in LaPt$_3$P. 
Ref.~\onlinecite{kang_cm}
found no indication of CDW instability, and supported a conventional 
electron-phonon ($EP$) scenario.

In this paper, we employ first principles calculations and
Migdal-Eliashberg theory to study superconductivity in the
$A$Pt$_3$P phosphides.
We find that SOC plays a marginal role in all three compounds and show
that the available experimental data are quantitatively
reproduced by conventional $EP$ theory,
based on well-converged electronic and phonon spectra.
This rules out exotic pairing and
CDW instabilities.\cite{chen12}
In fact, the $A$Pt$_3$P series is a textbook example for
$EP$ superconductivity: LaPt$_3$P, where $T_c$ is only 1.5 K, is a
typical low-$T_c$ superconductor in which all phonon branches are
moderately coupled to the electrons at $E_F$.
Lowering the electronic filling from trivalent La to divalent Sr and
Ca brings about an intense $EP$ coupling that is concentrated in
low-lying phonon branches with substantial Pt in-plane breathing
character.
In SrPt$_3$P, these branches are flat and have low frequencies,
and this entirely explains the large value of its BCS
ratios,\cite{taka12,marsiglio_c} with no need for multiband effects.

\begin{figure}
  \includegraphics[width=0.8\columnwidth]{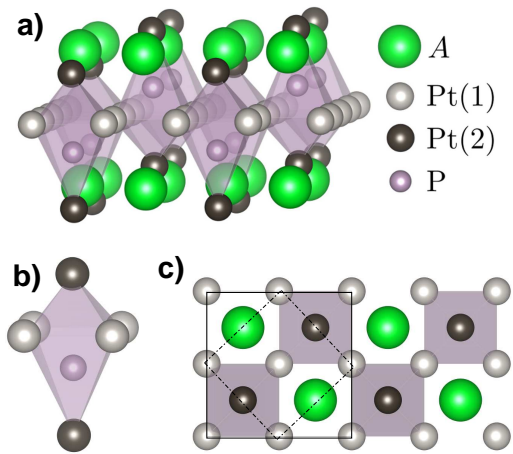}
  \caption{(Color online) ($a$) Crystal structure of $A$Pt$_3$P, space
    group $P$4/$nmm$.  The $\sqrt{2} \times \sqrt{2}$ in-plane
    arrangement of the distorted $X$Pt$_6$ octahedra ($b$)
    distinguishes the $A$Pt$_3$P compounds from the
    non-centrosymmetric CePt$_3$Si superconductor (space group
    $P4mm$); the corresponding unit cells are shown as full and dashed
    lines in panel ($c$).}
  \label{fig:fig1}
\end{figure}

\section{Methods}

Our calculations employ density functional perturbation
theory,\cite{dfpt} within the generalized gradient approximation (GGA)
as implemented in the Quantum-ESPRESSO package.\cite{qe}  We use
ultrasoft pseudopotentials~\cite{ultra} and basis set cutoffs of 40 Ry
and 400 Ry for wave function and charge density, respectively. We use
an $8 \times 8 \times 8$ grid for zone integration in the
self-consistent calculations, while a denser $16 \times 16 \times 16$
grid is used in the electron-phonon coupling calculations. The
dynamical matrices are calculated on an $8 \times 8 \times 8$ grid,
and phonon dispersions and DOS are then obtained by Fourier
interpolation.
The results of the structural relaxation were tested with the 
all-electron code Wien2k, which employs the
full potential linear augmented plane wave method.\cite{wien2k,LAPW}

\section{Structure}


Fig.~\ref{fig:fig1}$(a)$ shows the crystal structure 
of the $A$Pt$_3$P phosphides (space group $P4/nmm$).
%
In these tetragonal antiperovskites, the ionic radii of
Pt and P ions are similar, and the P ions off-center from
the octahedral basal plane formed by the Pt(1) ions as the P
ion is too big to fit into the center of the Pt(1) square. The apical
Pt(2) ion that is further from the P ion also moves closer to the basal
plane to ensure closer packing of the constituent 
ions. 
The distortion of the $X$Pt$_6$ octahedra occurs in both the
phosphides ($X$=P) and in the non-centrosymmetric rare-earth silicides
CePt$_3$Si and LaPt$_3$Si ($X$=Si), with space group $P4mm$.  What
distinguishes the two structures is the in-plane arrangement of the
octahedra. In the silicides, the distortions have a \textit{polar}
arrangement, \textit{i.e.}\ they all point in the same direction, and
the resulting structure has no inversion symmetry.  The corresponding
unit cell that contains one formula unit (\textit{f.u.})  is shown
with dashed lines in Fig.~\ref{fig:fig1}$(c)$.  With full lines we
show the $\sqrt{2}\times \sqrt{2}$ cell of the phosphides, where the
distortions alternate in a checkerboard fashion and restore the
inversion symmetry (\textit{antipolar} structure). Therefore, the unit
cell comprises two \textit{f.u.}, and $A$, Pt(1), Pt(2) and P
occupy 2$a$, 4$e$, 2$c$ and 2$c$ Wyckoff positions respectively.  We
relaxed the structures fully within GGA such that the force on each
atom is less than $10^{-5}$ Ry/Bohr. The relaxed parameters are given
in Table \ref{tab:latt}.

\begin{table}[h!tbp]
  \caption{\label{tab:latt} Fully relaxed structural parameters (GGA)
    for the $A$Pt$_3$P compounds in the experimental $P$4/$nmm$
    structure.  }
  \begin{ruledtabular}
    \begin{tabular}{lcccc}
                & $a (\AA)$ & $c (\AA)$    &  $z_{Pt2}$ & $z_{P}$ \\
      \hline
      SrPt$_3$P & 5.898     & 5.470        & 0.1362    & 0.7227 \\
      CaPt$_3$P & 5.758     & 5.494        & 0.1357    & 0.7303 \\
      LaPt$_3$P & 5.838     & 5.553        & 0.1418    & 0.7719 \\
    \end{tabular}
  \end{ruledtabular}
\end{table}
\begin{figure}
  \includegraphics*[width=0.6\columnwidth]{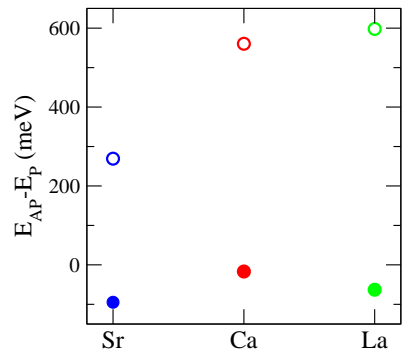}
  \caption{Total energy difference between the antipolar and polar
    structure for phosphides (full symbols) and selenides (empty
    symbols), in the fully relaxed GGA structure at zero pressure.}
  \label{fig:etot}
\end{figure}

The crystal structure of $A$Pt$_3$P, with space group $P4/nmm$, and
that of $RE$Pt$_3$Si ($RE$=La,Ce), with space group $P4/mm$, can be
seen as variants of the same antiperovskite structure, which differ
only for the in-plane polar arrangement of the distorted $X$Pt$_6$
octahedra. 
%
%
For $A$=La both the non-centrosymmetric ($P4/nmm$) silicide and the
centrosymmetric ($P4/mm$) phosphide compound exist, and are
superconducting with low $T_c$'s $\lesssim 2 K$, while for $A$=Ca, Sr
to our knowledge only the phosphides have been synthesized.

We have studied the relative stability of the two structures for
$A$Pt$_3$ silicides and phosphides, using total energy calculations.
The structures were fully relaxed in GGA at zero pressure. For the
existing compounds, the relaxed parameters are within $\sim$2$\%$ of
the experimental values.

\begin{figure*}
  \centering \includegraphics[width=2\columnwidth]{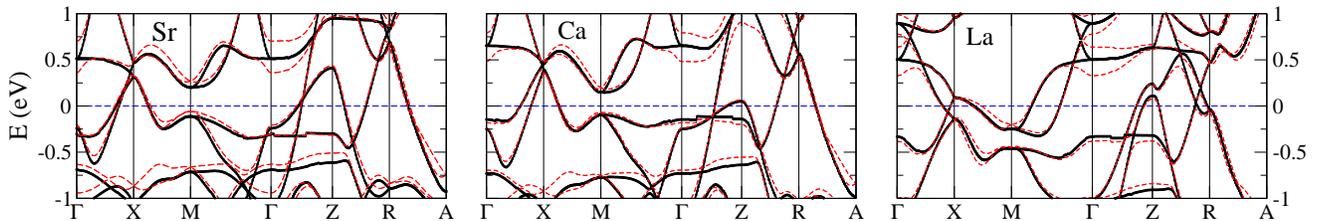}
  \caption{(Color online) Electronic structure of $A$Pt$_3$P with
    (red, dashed line) and without (black, solid line) spin orbit
    coupling (SOC); the zero of the energy is the Fermi level.}
  \label{fig:fig2}
\end{figure*}

\begin{figure*}
  \centering \includegraphics[width=2\columnwidth]{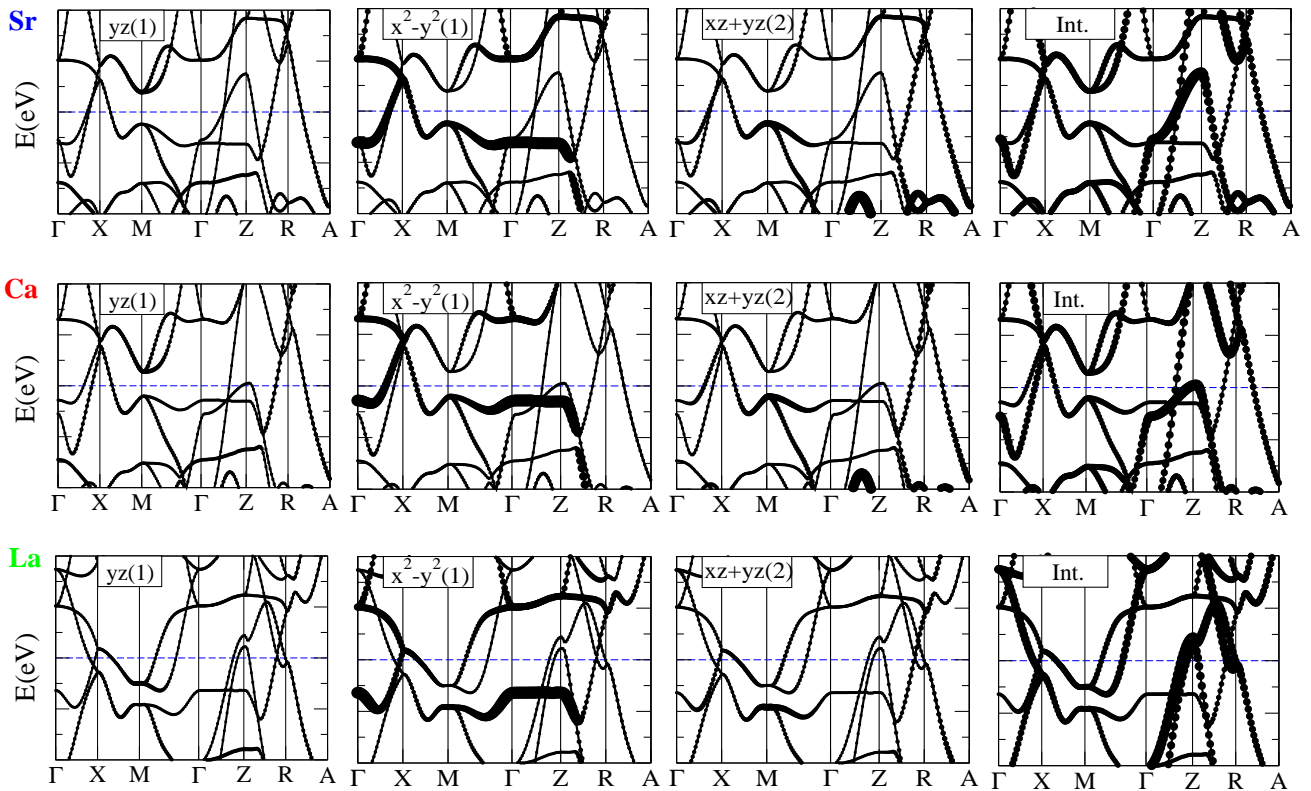}
  \caption{``Fat bands'' of $A$Pt$_3$P, decorated with partial orbital
    characters. $xy(1)$ and $yz(1)$ are in-plane Pt (Pt(1)) partial
    characters; $yz(2)$ refers to Pt apical atoms (Pt(2)); Int is
    interstitial.}
  \label{fig:fat}
\end{figure*}

In Fig.~\ref{fig:etot}, we plot the difference in total energies
between the antipolar and polar structures ($E_{AP} - E_{P}$).  The
difference is negative for phosphides (full symbols), and positive for
silicides (empty symbols), with approximately an order of magnitude
difference between the two cases.  The small energy differences found
in the phosphides (a few tens meV), and in particular for CaPt$_3$P,
indicates that it might be indeed possible to synthesize these
compounds also in the \textit{polar} variant, realizing the proposal
of Takayama \textit{et al.}\ to study the lack of inversion symmetry
in electronically-equivalent compounds.\cite{taka12}


\section{Electronic structure}

In the following, we stick to the $A$Pt$_3$P compounds in the $P4/nmm$
structure, employing fully relaxed lattice constants and internal
parameters.
Our electronic structures are in very good agreement with previous
calculations.\cite{chen12,nekr12} A zoom-in of the bands for energies
$\pm 1$ eV around the Fermi level ($E_F$), with (red, dashed) and
without (black, solid) SOC, is shown in Fig.~\ref{fig:fig2}. Only in
LaPt$_3$P does the SOC lead to a visible lifting of some band
degeneracies. But even in this case, the bands remain spin degenerate,
in contrast to what is claimed by Ref.~\onlinecite{chen12}. Since the
SOC plays only a marginal role on the electronic states near $E_F$, we
neglect it in the following.

Fig.~\ref{fig:fat} shows the ``fat'' bands without the SOC of
the three $A$Pt$_3$P compounds, highlighting the orbital characters
which give the highest contribution to the electronic states at the
Fermi surface.  The axes are oriented along the shortest in-plane
Pt-Pt distance.
%
%
The Fermi surface of the  $A^{2+}$ compounds comprises a large, low
dispersive sheet, formed by the band with prominent  Pt(1) $d_{x^2-y^2}$
character which crosses $E_F$ along the $\Gamma-X-M$ line; 
two other bands, with more Pt(2), P and interstitial character, 
form two more dispersive pockets,
\textit{i.e.} a large, flat, structure centered around the $\Gamma$ point,
and a small cigar-shaped hole pocket around the $Z$ point.
These bands cross $E_F$ along the $Z-R-A$ line.
This unequal distribution of orbital characters on the Fermi surface
suggests that the superconducting gap may be anisotropic.
%
%
LaPt$_3$P, with one more electron per $f.u.$ has its Fermi level
shifted up by $\sim$0.5 eV, and the band with low dispersion along
$M-\Gamma$ is completely full. The Fermi surface is composed of highly
dispersive sheets, with strongly mixed orbital contributions of Pt(2),
P and Pt(1) characters.

\section{Phonon dispersions and electron-phonon coupling}

\begin{figure}[h!ptb]
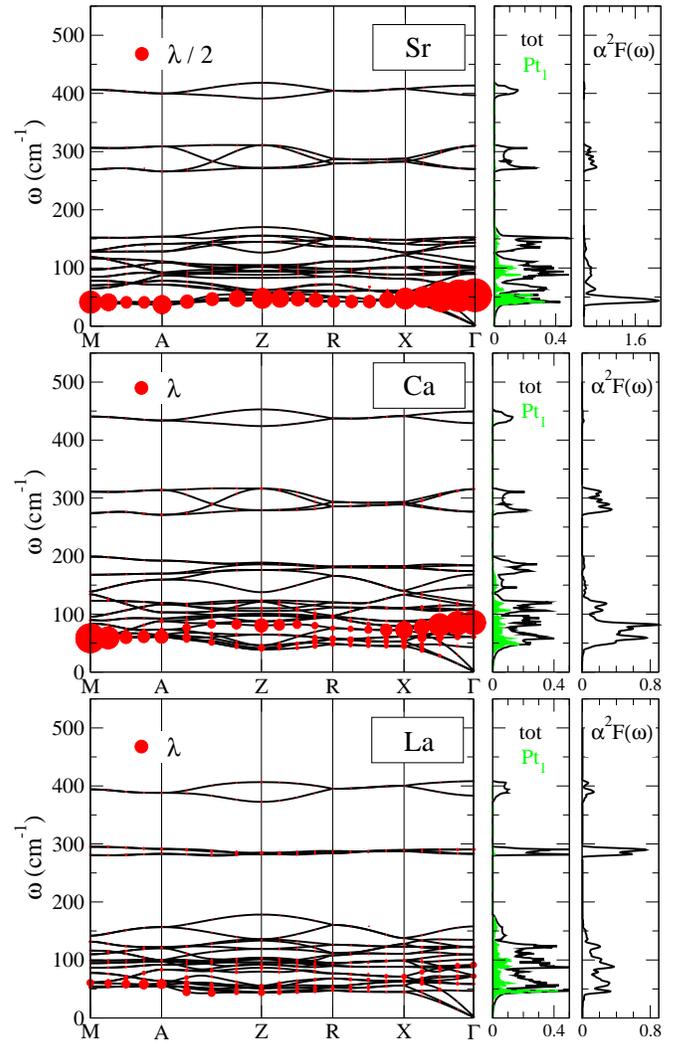

\centering
  \includegraphics*[width=\columnwidth]{FIG3a.eps}
  \includegraphics*[width=\columnwidth]{FIG3b.eps}
  \includegraphics*[width=\columnwidth]{FIG3c.eps}
  \caption{ (Color online) Phonon dispersions, density of states and
    Eliashberg function $\alpha^2 F(\omega)$ of $A$Pt$_3$P.  The
    phonon dispersions are decorated with symbols, proportional to the
    partial $EP$ coupling $\lambda_{\mathbf{q} \nu}$; for readability,
    the $\lambda$'s for Sr have been rescaled by a factor two. The
    logarithmically averaged phonon frequencies $\omega_{\ln}$, the
    $EP$ coupling constants $\lambda$, and the corresponding critical
    temperatures $T_c$'s are given in Table~\ref{tab:tc}.}
  \label{fig:fig3}
\end{figure}

The three compounds have very similar phonon dispersions, shown in the
left panels of Fig.~\ref{fig:fig3}. The thirty phonon branches extend
up to $\sim$450 cm$^{-1}$, with two upper branches of mostly
out-of-plane P character, four intermediate branches at $\sim$300
cm$^{-1}$ that show mostly in-plane vibration of P, and a lower
manifold of the twenty-four strongly intertwined branches, with mixed
$A$, Pt(1) and Pt(2) character.

There is a substantial difference in the three Eliashberg
functions $\alpha^2 F(\omega)$ , plotted in the rightmost panels of
Fig.~\ref{fig:fig3} and defined as: 
\begin{equation}
\alpha^2 F(\omega)=\frac{1}{N(0)}\sum_{\mathbf{k},\mathbf{q},\nu,n,m}%
\delta(\epsilon_{\mathbf{k}}^{n})\delta(\epsilon_{\mathbf{k+q}}^{m}%
)|g_{\mathbf{k},\mathbf{k+q}}^{\nu,n,m}|^{2}\delta(\omega-\omega^{\nu}
_{\mathbf{q}}),
\label{eq:alpha}
\end{equation}
where $\omega^{\nu}_{\mathbf{q}}$ are phonon frequencies,
$\epsilon_{\mathbf{k}}^{n}$ electronic energies, 
and
$g_{\mathbf{k},\mathbf{k+q}}^{\nu,n,m}$  $EP$ matrix
elements.
%
%
The $\alpha^2 F(\omega)$ yield information not only on the intensity of the
total $EP$ coupling ($EPC$) $\lambda$, but also on the nature of the
bonding and character of the superconducting state.
In general, an $\alpha^2 F(\omega)$
roughly proportional to
the PDOS is characteristic of metals with a weak to
moderate total coupling and low $T_c's \lesssim 5$ K.
The best $EP$ superconductors,
such as MgB$_2$ and A15's, are instead characterized by
 $\alpha^2 F(\omega)$ which display sharp peaks only at
specific parts of the phonon spectrum,
 reflecting a strong coupling between specific electron and phonon states. This 
requires (partly) covalent bonding.
It is not uncommon that within the same family of
materials the electron count changes the $EP$ coupling regime from
weak to strong, depending on the nature of the electronic states at
$E_F$ selected by  the two $\delta$ functions in Eq.~\ref{eq:alpha}.

Fig.~\ref{fig:fig3} shows that while the coupling
is uniform in LaPt$_3$P, in the two $A^{2+}$ compounds it is strongly enhanced 
at low frequencies.
Furthermore, it is almost entirely concentrated in the low-lying
phonon branches with substantial Pt(1) in-plane breathing character,
which at the $\Gamma$ point has $B_{2u}$ symmetry --- $\omega_{\mathbf{q}}^{\textrm{br}}$ in the
following.\cite{chen12,kang_cm}
These modes are very low in energy and
almost dispersionless in SrPt$_3$P ($\omega \sim 50$ cm$^{-1}$) and
slightly harder in Ca, where they have a sizable dispersion that extends up to $\sim$100
cm$^{-1}$.
One can trace their evolution in the two compounds by
following the largest $\lambda^{\nu}_{\mathbf{q}}$ symbols that decorate the
phonon dispersions, or looking at the partial Pt(1) phonon DOS 
plotted in the middle panel of Fig.~\ref{fig:fig3}.
In-plane Pt breathing modes couple more strongly to the  Pt(1) in-plane
electronic states,
and less to other partial characters.
This causes some anisotropy in the $\mathbf{k}$
space distribution of the $EPC$ and, as a consequence, in the 
$\mathbf{q}$-dependence of the $\lambda^{\nu}_{\mathbf{q}}$
in the  $A^{2+}$Pt$_3$P, and explains the much lower coupling in  LaPt$_3$P.

The total $EPC$ constant $\lambda\!=\!\sum_{\mathbf{q},\nu} 
\lambda_{\mathbf{q}}^{\nu}=2 \int_{0}^{\infty} 
\frac{\alpha^2 F(\omega)}{\omega} d \omega$
is quite low in LaPt$_3$P ($\lambda=0.57$) but sizable both in
CaPt$_3$P ($\lambda=0.85$) and SrPt$_3$P ($\lambda=1.33$).
Due to the $1/\omega$ factor, the $EPC$ is strongly enhanced in
SrPt$_3$P with respect to CaPt$_3$P because a considerable part of the
breathing branches is shifted to low frequencies.  A similar softening
of the breathing branch is discussed by Chen \textit{et al.},\cite{chen12} who
find a dynamical instability of the breathing branch, which we cannot
reproduce.\cite{foot:chen} For some $\mathbf{q}$ points where the
differences in $\omega_{\mathbf{q}}^{\textrm{br}}$ are large, the
partial $EPC$ constants $\lambda_{\mathbf{q}}^{\textrm{br}}=const
\times
I_{\mathbf{q}}^{\textrm{br}}/(\omega_{\mathbf{q}}^{\textrm{br}})^2$
differ by as much as a factor three despite a very small difference
($\sim\!10 \%$) in the matrix elements ($I^{\textrm{br}}$), as shown
in Table \ref{tab:comp} for the $\Gamma$ point.

\begin{figure}
  \includegraphics*[width=0.8\columnwidth]{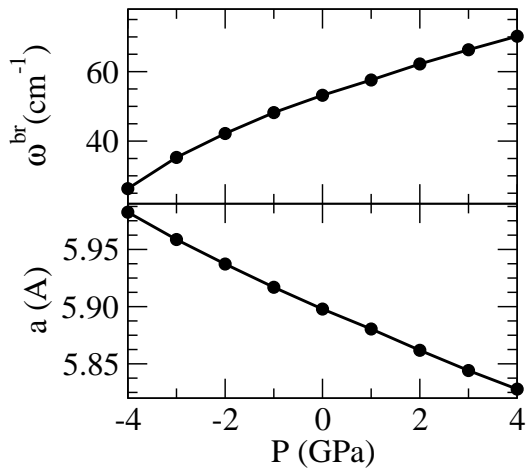}
  \caption{SrPt$_3$P: Calculated frequency of the B$_{2u}$ mode of
    SrPt$_3$P (top) and in-plane lattice constant (bottom) as a
    function of pressure.  }
  \label{fig:freq}
\end{figure}

The small difference in the EP matrix elements implies that the
lowering of the frequencies in Sr with respect to Ca is not due to
increased EP coupling.
Rather, this is almost entirely a structural effect, since we find
that $\omega_{\mathbf{q}}^{\textrm{br}}$ is increased (reduced) by reducing (increasing) the in-plane lattice constant, which in CaPt$_3$P is $\sim$0.14 \AA smaller than in SrPt$_3$P.
To confirm the strong sensitivity of the Pt(1) breathing branch to the
in-plane lattice constant, we have calculated the $\Gamma$-point
phonon frequencies of SrPt$_3$P as a function of pressure using the
theoretical structures that are optimized within the GGA. The
frequency of the B$_{2u}$ mode as a function of pressure is plotted in
Fig.~\ref{fig:freq} together with the corresponding value of the
in-plane lattice constant $a$. For pressures of $\pm 4$ GPa, we estimate a linear
dependence $\omega^{\textrm{br}}(P)= + 5 cm^{-1} /GPa$. This
translates in an almost linear dependence in terms of the in-plane
lattice constant: $\omega^{\textrm{br}}(a) \sim -20$ cm$^{-1}$/\AA.
From the calculated pressure derivative of the zone-center frequency
of SrPt$_3$P, we estimate that this would correspond to a 30 cm$^{-1}$
hardening of the phonon frequency, in remarkable agreement with the
values calculated for CaPt$_3$P.

\begin{table}[h!tbp]
  \caption{\label{tab:comp} Some calculated results for
    $A$Pt$_3$P: Electronic density of states at the Fermi level
    $N(0)$ in states/eV/spin/unit cell; $\omega_{\ln}$ 
    in $K$; Frequency 
    $\omega^{\textrm{br}}$ in cm$^{-1}$, e-ph coupling
    $\lambda^{\textrm{br}}$, and e-ph coupling strength
    $I^{\textrm{br}}$ in cm$^{-2}$ for the in-plane
    Pt(1) breathing mode at $\Gamma$.
  }
  \begin{ruledtabular}
    \begin{tabular}{lcccccc}
                & $N(0)$     & $\lambda$    &  $\omega_{\ln}$ & $\omega^{\textrm{br}}$ & $\lambda^{\textrm{br}}$ & $I^{\textrm{br}}$ \\
      \hline
      SrPt$_3$P & 2.36       & 1.33         &  77             &       53.0             & 1.90                    &   5337           \\  
      CaPt$_3$P & 2.37       & 0.85         &  110            &       85.5             & 0.63                    &   4605           \\
      LaPt$_3$P & 1.94       & 0.57         &  118            &       91.5             & 0.14                    &   1172           \\
    \end{tabular}
  \end{ruledtabular}
\end{table}

The sensitivity of $\omega_{\mathbf{q}}^{\textrm{br}}$ to the value of
the in-plane lattice constant explains why the regime for
superconductivity is weak coupling in Ca and strong coupling in Sr,
despite the very close critical temperatures. In fact, the shift of
spectral weight in $\alpha^2 F(\omega)$ to lower energies causes a
strong enhancement in $\lambda$, but it also induces a decrease in
$\omega_{\ln}=\exp\left(\frac{2}{\lambda}\int_0^{\infty}
d\omega/\omega \alpha^2 F(\omega) \ln \omega \right)$---see Table
~\ref{tab:comp}.  These two factors compensate in the Allen-Dynes
expression for T$_c$:
\[
T_c= \frac{\omega_{\ln}}{1.20}\exp\left( - \frac{1.04(1+\lambda)}{\lambda-\mu^*-0.62\lambda\mu^*}\right),
\] 
but not in the BCS ratios ($2\Delta/T_c$, $\Delta C/T_c$, etc.), which
to a very good approximation depend only on the quantity
$T_c/\omega_{\ln}$.

\section{Migdal-Eliashberg Theory}

\begin{figure}
  \includegraphics*[width=0.75\columnwidth]{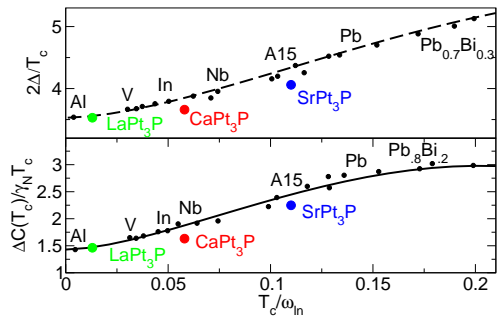}
  \caption{(Color online) Location of the $A$Pt$_3$P compounds on the
    Marsiglio-Carbotte plots for strong-coupling superconductors.
    $\Delta$ is the superconducting gap at zero temperature, $T_c$ is
    the critical temperature, $\Delta C (T_c)$ is the jump in the
    electronic specific heat at $T_c$, $\gamma$ is the linear
    coefficient of the normal-state specific heat, obtained from DFT
    and single-band Migdal-Eliashberg theory and $\omega_{ln}$ is the
    logarithmic averaged phonon frequency (see text).  Lines are
    obtained from approximate solution of the Migdal-Eliashberg
    equations.  Figures are adapted from
    Ref.~\onlinecite{marsiglio_c}.  In increasing order the black
    points correspond to the following systems: Al, V, Ta, Sn, Tl, In,
    Nb (Butler), Nb (Arnold), Nb (Robinson), Nb$_{0.75}$Zr$_{0.25}$,
    V$_{3}$Ga,Nb$_3$Al, Nb$_3$Ge, Pb, Pb$_{0.8}$Tl$_{0.2}$,
    Pb$_{0.9}$Bi$_{0.1}$, Pb$_{0.8}$Bi$_{0.2}$, Pb$_{0.7}$Bi$_{0.3}$
    and Pb$_{0.65}$Bi$_{0.35}$.  Data taken from
    Ref.~\onlinecite{marsiglio_c} and references cited therein.}
\label{fig:mc}
\end{figure}

In Ref.~\onlinecite{marsiglio_c} Marsiglio and Carbotte have shown
that the BCS ratios of all known superconductors fall on a universal
curve when plotted as a function of $T_c/\omega_{\ln}$.
We have obtained the values of the superconducting and thermodynamical
quantities from the full solution of the single-band Migdal-Eliashberg
equations to locate the $A$Pt$_3$P compounds on the Marsiglio-Carbotte
plots. As shown in Fig.~\ref{fig:mc} and from the data summarized in
Table.~\ref{tab:tc}, LaPt$_3$P and CaPt$_3$P, with
$T_c/\omega_{\ln}$=0.013 and 0.058, respectively, lie together with
elemental metals, while SrPt$_3$P ($T_c/\omega_{\ln}= 0.110$) is
placed at the lower end of a broad class of the low-phonon,
strong-coupling superconductors, together with the A15 and Chevrel
compounds.

In order to calculate the critical temperatures, gap values and
specific heat data presented in Fig.~\ref{fig:mc} and Table
\ref{tab:tc} we solved the Migdal-Eliashberg
equations~\cite{M-E-review} in the single-band case:
\begin{eqnarray*}
  \phi(\omega_n)  = \pi
  T\sum_{m=-M}^{m=M}[\lambda(\omega_n-\omega_m)-\mu^*]\frac{\phi(\omega_m)}{\sqrt{\omega_m^2Z^2(\omega_m)+\phi^2(\omega_m)}}
\end{eqnarray*}
\begin{eqnarray*}
  Z(\omega_n)\omega_n  = \omega_n +\pi
  T\sum_{m=-M}^{m=M}\lambda(\omega_n-\omega_m)\frac{Z(\omega_m)\omega_m}{\sqrt{\omega_m^2Z^2(\omega_m)+\phi^2(\omega_m)}}
\end{eqnarray*}
\begin{eqnarray*}
  \lambda(\omega_n-\omega_m) & = & 2\int_0^{\infty}\frac{\Omega\alpha^2F(\Omega) d \Omega}{(\omega_n-\omega_m)^2+\Omega^2},
\end{eqnarray*}
where $\phi(\omega_n)=\Delta(\omega_n)Z(\omega_n)$, $\Delta(\omega_n)$
is the superconducting gap, $Z(\omega_n)$ is the mass enhancement
factor and $M$ is the number of Matsubara frequencies $\omega_n$ used
in the calculations.  We used the Eliashberg functions in
Fig.~\ref{fig:fig3} and the densities of states in Table
\ref{tab:comp}.  The value of $\mu^*$ was chosen to reproduce the
experimental $T_c$---we obtained $\mu^*=0.1 \pm 10 \%$ for all three
compounds---and kept fixed in the calculations of the specific heat
jump and superconducting gap.
The specific heat jump was obtained by numerically calculating the
difference $\Delta F(T)$ between the normal state (N) free energy and
the superconducting one (S):~\cite{Dolgov}
\begin{eqnarray*}
  \Delta F(T) &=& -\pi T \sum_{m=-M}^{m=M}\left \{
    | \omega_n|(Z_N(\omega_n)-1) \right.\\
 &-&\frac{2\omega_n^2 [Z^2_S(\omega_n)-1]+2\phi^2(\omega_n) }{|\omega_n|+\sqrt{\omega_n^2Z_S^2(\omega_n)+\phi^2(\omega_n)}} 
\\
&+&\left. \frac{\omega_n^2Z_S(\omega_n)(Z_S(\omega_n-1))+\phi^2(\omega_n)}{\sqrt{\omega_n^2Z_S^2(\omega_n)+\phi^2(\omega_n)}}
  \right \}
  \label{eq:deltaF}
\end{eqnarray*}
and fitting the obtained curve with a  12$^{th}$ order polynomial. 
The specific heat was then obtained from the second derivative of the polynomial expansion.

\begin{figure}[h!tbp]
\centering
  \includegraphics*[width=6cm]{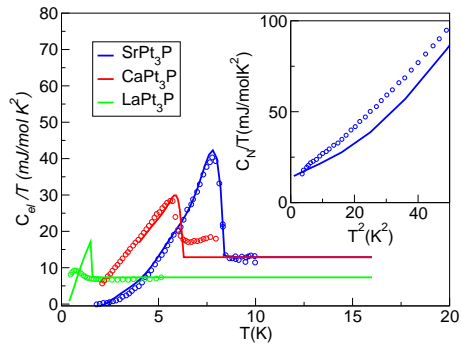}
  \caption{\footnotesize{
(Color online) Comparison between experimental data from
Ref.~\onlinecite{taka12} (colored dots) and 
Migdal Eliashberg theory (colored lines) for the heat capacities
of $A$Pt$_3$P ($A=$ Sr, Ca and La).
Inset: Comparison between experimental data from
Ref.~\onlinecite{taka12} (blue dots) and first-principles
calculations (blue line) for the normal state specific heat
of SrPt$_3$P.}}
  \label{fig:heat}
\end{figure}

Fig.~\ref{fig:heat} shows a comparison of the specific heat data with
experiments from Ref.~\onlinecite{taka12}.  In the inset, we compare
calculations (solid lines) and experiments (symbols) for the total
normal state specific heat of SrPt$_3$P.  The lattice specific heat is
fitted with: $c (T) = b T^3 + d T^5$, $b$ = $1.29 (1.26)$ mJ/mol
K$^4$, $d$=$8.9 (13.0)$ mJ/mol K$^6$ for theoretical (experimental)
data, respectively.
This remarkable agreement suggests that our calculated phonon spectra,
with a sharp peak at $\omega \sim 50$ cm$^{-1}$, are very close to the
actual ones. This allow us to rule out the dynamical instabilities
calculated in Ref.~\onlinecite{chen12}.

In the main panel of Fig.~\ref{fig:heat}, the superconducting state
data for SrPt$_3$P are shown as a blue curve. The single-band Migdal
Eliashberg calculations, which yield $\frac{2 \Delta}{T_c} \sim 4.06$,
reproduce almost exactly the experimental curve.
As for the low-$T_c$ compounds, we again obtain critical temperatures
in very good agreement with experiments, with $\mu^* = 0.1$, which is
a strong indication in favor of conventional superconductivity.
CaPt$_3$P (red curves in Fig.~\ref{fig:heat}) displays a mass enhancement in the normal state that is 2.5 times larger than the 
calculated one, which we attribute to the presence of additional superconducting
phases in the sample.
All other quantities are in line with our
calculations. In fact, using the calculated value of $\gamma_N=10.3$ mJ mol$^{-1}$ K$^{-2}$,
we obtain also a reasonable fit to the specific heat jump, as shown in Fig.~\ref{fig:heat}.
%
For LaPt$_3$P (green curves in Fig.~\ref{fig:heat}) 
we obtain a fairly good agreement for the normal state quantities,
while the superconducting state data are probably too noisy
for a meaningful comparison.

Fig.~\ref{fig:mc} clearly shows that the values we obtain for all
three $A$Pt$_3$P compounds are fully in line with other $EP$
superconductors.  The very high BCS ratio $2 \Delta/T_c=5.0$ reported
from Takayama \textit{et al.}\cite{taka12}\ for SrPt$_3$P lies instead out of
the general trend.  In Fig.~\ref{fig:mc}, $ 2 \Delta/T_c = 5.0$
corresponds to $T_c/\omega_{\ln} \sim 0.18$, which, given the
calculated $\omega_{\ln}$, leads to a $T_c \sim 14$ and $\lambda\sim
3$, clearly inconsistent with the experiment.
%
%
In Ref.~\onlinecite{taka12} the value $2 \Delta/T_c=5.0$ is one of the
strongest arguments for multiband superconductivity. We believe that
this is an artifact of the simplified $\alpha$ model used by the
authors to fit the experimental specific heat.  In fact, as we have
shown above, a single gap Migdal Eliashberg model perfectly fits the
electronic specific heat for SrPt$_3$P, with a lower $2\Delta/T_c =
4.06$. Therefore, also the possible anisotropy in the gap suggested by
the uneven distribution of orbital characters on the Fermi surface is
either negligible, or washed out by impurities in real samples.  At
the same time, the good agreement of the calculated lattice specific
heat, allows us to rule out the dynamical instability of the low-lying
phonon branches in SrPt$_3$P and the CDW scenario based
thereon.\cite{chen12}
%
%
%
%
%
The same single-band analysis, applied to the lower $T_c$ compounds,
yields critical temperatures in very good agreement with experiments,
which is a strong indication in favor of conventional
superconductivity.

%
%
%
\begin{table}[h!tbp]
  \caption{\label{tab:tc} Superconducting properties of $A$Pt$_3$P, from
   first-principles calculations and Migdal-Eliashberg theory;
  $\gamma_N$ is the electronic normal-state specific heat, 
 in mJ mol$^{-1}$ $K^{-2}$,
 $\Delta (0)$ is the value of the superconducting gap, $\Delta C$ is the specific heat 
 jump at $T_c$. Experimental data from Ref.~\onlinecite{taka12} are in
 parentheses. 
The Coulomb pseudopotential $\mu^*$ was fixed to reproduce the experimental
$T_c$.
 }
  \begin{ruledtabular}
    \begin{tabular}{lcccccc}
         &     $\gamma_N$       &   $T_c$       &   $2\Delta(0)/T_c$  &
         $\Delta$ $C/T_c$   & $T_c/\omega_{\ln}$ & $\mu^*$ \\
      \hline
      Sr &     12.9 (12.7)      &   8.5 (8.4)   &    4.06               &  29.0 (28 )        &   0.110   & 0.11         \\
      Ca &     10.3 (17.4)      &   6.34 (6.6)   &    3.66               &
      16.8 (11 )         &   0.058  & 0.09           \\ 
      La &     7.18 (6.7)       &   1.56 (1.5)   &     3.53              &
      10.5 (2 )         &   0.013  & 0.11          \\
    \end{tabular}
  \end{ruledtabular}
\end{table}

\section{Conclusions}
In conclusion, the first-principles calculations and Migdal-Eliashberg
analysis presented in this work allow us to make some conclusive
statements about the nature of superconductivity in the
recently-discovered $A$Pt$_3$Pt compounds
($A$=Sr, Ca, La).\cite{taka12} Superconductivity in $A$Pt$_3$Pt
($A$=Sr, Ca, La) compounds is of conventional $EP$ nature and the SOC
plays a negligible role, thus ruling out the proposals of exotic
superconductivity of Refs.~\cite{chen12,nekr12}.  The electronic
filling brings about an $EP$ coupling which is moderate in La, and
much stronger in Ca and Sr, where Pt(1) breathing phonons couple to
in-plane electronic states.  The frequency and dispersion of the
breathing phonons can be \textit{tuned} acting on the in-plane lattice
constant, leading to weak and strong-coupling values of the BCS ratios
in Ca and Sr, despite the very close critical temperatures.

Furthermore, our total energy calculations suggest 
that the $A$Pt$_3$P compounds could also be synthesized
in the related, non-centrosymmetric CePt$_3$Si structure,
through appropriate synthesis
conditions or partial replacement of P with Si.
This would realize the original proposal of Takayama \textit{et al.},\cite{taka12}
and open the way to the exciting possibility of studying 
the effect of the lack of inversion symmetry on superconductivity
in a \textit{controlled} way.

\section{Acknowledgements} 
We would like to acknowledge A. P. Schnyder, O.V. Dolgov and
R.K. Kremer for useful discussions.




\begin{thebibliography}{99}
\bibitem{foot:chen}
We carefully checked the convergence of our phonon spectra, 
in particular at the $X$-point, but found no indication of instabilities. The
difference most likely stems from the different pseudopotentials employed in
their work. 
\end{thebibliography}
\end{document}